\documentclass[11pt]{elsarticle}
\usepackage{amsmath,amssymb}
\usepackage[margin=2.5cm]{geometry}
\usepackage{graphics,graphicx}
\usepackage{dcolumn,bm}
\usepackage{psfrag}
\usepackage{color}
\newcommand{\be}{\begin{equation}}
\newcommand{\ee}{\end{equation}}

\newcommand{\epsi}{\epsilon}
\usepackage{graphics,graphicx}
\journal{Physica A}

\begin{document}
\title{Block size dependence of coarse graining in discrete opinion dynamics model: Application to the US presidential elections}

\author{Kathakali Biswas${}^1$}
\author{Soumyajyoti Biswas${}^{2}$}
%\email{soumyajyoti.b@srmap.edu.in}
\author{Parongama Sen${}^{1}$}
\address{${}^1$Department of Physics, University of Calcutta, 92 Acharya Prafulla Chandra Road, Kolkata 700009, India.\\
{${}^2$Department of Physics, SRM University - AP, Andhra Pradesh - 522502, India}}

\date{\today}

\begin{abstract}
The electoral college of voting system  for the US presidential election is analogous  
 to a coarse graining procedure commonly used to study phase transitions in physical systems.  
 In a recent paper, opinion dynamics models manifesting a phase transition, were shown to 
 be able to explain the cases when a candidate winning more number of popular votes could still 
lose the general election on the basis of the electoral college system. 
We explore the dependence of such possibilities on various factors 
like the number of states and  total population (i.e., system sizes)  and get an 
interesting scaling behavior. In comparison with the real data, it is shown that the probability of the minority win,
calculated within the model assumptions, 
is indeed near the highest possible value.  In addition, we also 
implement a two step  coarse graining procedure, relevant for both  opinion dynamics  and  information theory.

\end{abstract}

\maketitle

%%%%%%%%%%%%%%%%%%%%%%%%%%%%%%%%%%%%%%%%%%%

\section{Introduction}
The concept that human society with a large number of mutually interacting agents can be regarded as a complex 
system consisting of an ensemble of basic units following simple rules,  has led to the investigation of several 
social phenomena using models commonly studied in statistical physics.
One of  the most studied social phenomena has been the formation of  opinion,
 where 
interaction between agents are considered at a microscopic level in the models, 
with or without external fields. The aim is often to find out the mechanism 
by which it is possible to achieve an agreement or consensus. 
%however,  
%The dynamical  evolution of opinion formation in a society is a complex process involving myriads of socio-economic,
%not to mention psychological, issues. 
The first step in such modeling approaches is to quantify the opinions of the agents involved. This task is easier 
 \cite{stauffer,sen_chak,soc_rmp,galam_book} when the
choice of the opinion is binary. Examples of such cases are    yes/no referenda, a two party voting etc., where the said quantification is straight-forward
with the use of a binary variable to represent the support for either side.  It is
often done with $\pm 1$ values, i.e., like spins with Ising symmetry, for example in the voter model \cite{cliff,ligg1,ligg2}. Depending on the situation,  it may also  be crucial to have
neutral opinions - which can be assigned a  zero value - representing the population that does not support either of the two groups or  sides \cite{bcs_us}.   
Then by including suitable interactions among the agents the dynamics of opinion formation, particularly the formation of a majority on either side through a symmetry breaking transition, are studied. 

The formation of the majority opinion is very important as it often determines the final outcome of the election process (for example, the question of the UK leaving the EU \cite{bcs_brexit}). However, in situations where the direct majority is neither the necessary nor the sufficient condition for winning the election, the spatial fluctuations in the voting 
patterns play a crucial role. Indeed, in hierarchical systems of voting, political paradoxes such as the overall winner not representing the majority opinion, can occur \cite{sg1,sg2,erikson}. Particularly,   
in this paper, we are concerned about the voting process of the US 
president-ship \cite{election_book}.  This is not a direct election process for the post of the president; an intermediate body, namely the electoral college delegates, 
 ultimately  decide who will be  the president. In the electoral college system,   
in most cases, all the  delegates from a particular state are assigned to the winner from that state and the overall winning candidate  is then decided on the basis of the support of the majority of those delegates. This process has led to the rather rare cases when the popular candidate
(PC henceforth), i.e., the one with the maximum number of individual votes could turn out to be the loser after the electoral college system is applied. Clearly, this is a result of strong local fluctuations in the voting pattern of the different states and more likely to happen in closely fought elections.
If there are two candidates only, say A and B,  the voters may be assigned the state values $\pm1$ to express their support, e.g, state 1 means supporting A, and in a closely fought election the  opinions values 1 and -1 are almost equal
in number.
A  closely fought election  with two candidates and the strong fluctuations together   resemble the behavior of physical systems (e.g., magnetic systems represented by the Ising model)  near their critical point. Indeed, some of the present authors had shown, using a suitable model dynamics, that close to the critical point of the model, a finite probability exists for the minority win situation, which becomes vanishingly small in the parameter range of the actual voting data when a (spatially) completely uncorrelated random distribution of the opinions are considered \cite{bcs_us}.     

It was further argued in Ref. \cite{bcs_us} that the election process with the electoral college system is analogous to  a coarse graining 
procedure commonly used to study phase transitions in, for example, magnetic materials.  In reality, most of the US presidential elections involve two major candidates only.  
Thus the  scenario is similar to the Ising model, as mentioned earlier,  and in that language, the sign of the magnetization  (which is   
defined as sum over all the spins/opinions),  based on all
the individual votes decides the PC. 
Magnetization is  the order parameter which for a fully disordered system will 
vanish. In the case of a two party voting, its sign determines the winning candidate. The 
process of electoral college voting then corresponds to  a one step coarse graining procedure,  implemented by a block transformation \cite{skma},
where a block represents a state. If the 
magnetization before and after the coarse graining have the same sign, the PC  emerges as the global winner (GW henceforth) and becomes the President. 
In the opposite case, PC loses and such rare cases have been witnessed four times in the history of the US presidential elections, the last time it happened was in 2016. 
Such cases can happen if the election is closely fought, or in other words,  the magnetization (the order parameter in general)
is small before the block transformation. 
Hence, in order to capture this behavior, one has to consider   a model
manifesting a phase transition and apply  the coarse graining close to criticality. 

%One can ask quantitatively what do we mean by closely fought. 
%As mentioned before, a closely fought election would render almost equal number of $\pm 1$ states and which corresponds to a near critical situation
To simulate the closely fought election, 
the initial state can be considered to be random and the correlation is allowed to grow in time under the dynamical scheme.  The value of the 
parameter driving the phase transition in the model is  chosen to be close to criticality.
Due to  the large fluctuations near criticality, the order parameter oscillates
about zero, the  amplitude is finite in a finite sized  system.  The (small) difference in the votes therefore is a variable and the calculation is made for  
all values that occur  over time. 
%in a single realization 
%near criticality.  

%The opposite can happen for a totally random voting of course but 
%that is far from reality and it was shown that one has to consider 
%models manifesting a phase transition (order-disorder) very close to
%the critical point. For  details  
%we encourage the reader to refere to  \cite{}. 

Although two variables $\pm1$ are sufficient  to describe the scenario such that one can use the Ising model as a proper opinion dynamics model here, 
 the results are closer to  reality when 
one considers a third population with neutral opinion \cite{bcs_us}. This was possible by  
taking 
 a model having three opinion values, namely, the kinetic exchange model
proposed in \cite{bcs}. Note that the order parameter (magnetization) remains same even after including the neutral opinions. 

\begin{figure}
\includegraphics[width = 10cm]{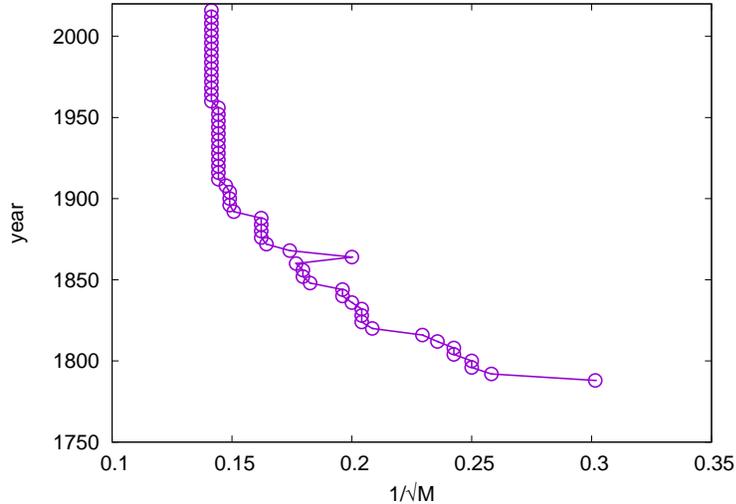}
\caption{In the US, the number of states ($M$) has changed over the years. As discussed in the text, while calculating the minority win probability, the relevant scaling variable
turns out to be $b/L$, which is $1/\sqrt{M}$ assuming uniform population density. The figure shows the variation of $1/\sqrt{M}$ with time in the US. The variation is non-monotonic in time during the period of the civil war, due to the confederate states. Other than that $1/\sqrt{M}$ varies from $1/\sqrt{11}$ in 1788 to $1/\sqrt{50}$ in 1959 and remained constant thereafter. }
\label{states} 
\end{figure}

 Specifically,  the parameter values were chosen to be as far as possible compatible to  the  actual case, e.g., the coarse graining was done 
by dividing the system  into  49  blocks (the numerical method required a square number)  close to  the present number of states in the US which is 50.   
However, the number of states in the US has varied considerably over time as shown in Fig. \ref{states}. Since the first ratification of admissions during 1787-88, 
the number of states in the US continued to increase, until the period of the civil war (1861-65), during which the confederate states were in existence, reducing the 
total number of states in the US. Since the end of the civil war, the total number of states kept increasing, until the present tally of 50 was reached in 1959 with the admission of Hawaii. 
  
In this work, we therefore generalize the problem and ask the question: how does the probability that the PC  turns out to be the  loser after the so called coarse graining, depend on the  number of states. We may thus predict 
 what would be the probability of a PC losing if some states are joined/split in future. 
We have  used the  Kinetic exchange model (KEM) and the Ising model (IM) (both show   an order disorder phase transition) on two dimensional square lattices for different system sizes $N = L^2$ and applied coarse graining with  different scale factors $b$. We note 
$b/L$  
emerges as the relevant scaling variable which is identical to  $\sqrt{1/M}$  where $M$ is the number of states. 
%quantity This means one can essentially 
%vary the number of states.  We have  also used different populations to check the dependence.   

The entire procedure can also be regarded as a problem in  information theory. Particularly, the time series of the sign of the opinion favoring the PC can be thought of as the input string and that of the GW the output string for the electoral college or the coarse graining process. The relative mutual information (see Eq. (4) in Ref. \cite{bcs_us}) is then a measure of the probability of faithful translation of the popular opinion by the electoral college, which is close to unity in the ordered state and decreases sharply near the critical point. 
 
Hence  the estimate in which we are interested, the probability that the coarse grained result of the order parameter is opposite in sign,  
 is generally termed the error $\epsi(b,L)$, studied as a function of $b$ and $L$. $b^2$, the area of a block is an integer, $b$ itself can be  irrational. For a given $L$, the maximum value of $b$ is $L/\sqrt{2}$.  
Two limiting cases are immediately identified. 
If $b$ is trivially 1, then each individual   represents a  state and essentially no  coarse graining is required
and $\epsi(1,L) =0$. 
  On the other hand, when $b = L$, we are 
taking the whole population as a block and the coarse graining will merely 
yield the same value of the order parameter  making $\epsi(L,L) =0$. Intermediate values of $b$ will show a point of maximum error i.e., the highest probability for the minority win. Given that the number of states in the US has varied over the years, it is interesting to identify the proximity of the $b/L$ value to the point where the maximum error is predicted from the model.

\section{Models and quantities calculated}

As mentioned earlier, we study the Ising model (IM) and a kinetic exchange model (KEM) in capturing the dynamics of essentially two-party voting systems. 
The Ising model represents a  binary opinion case with $s_i = \pm 1$, the opinion held by the $i$th agent at the site $i$. The two values are used to define the support for the two candidates.  
The Ising model with the Hamiltonian $H = -J\sum_{<ij>} s_is_j $ is studied close to the exactly known critical point using  Metropolis algorithm. 
Here $<ij>$ denotes nearest neighbors.

For the KEM, the opinion values can be $o_i = \pm 1$ and 0. 
For the  $i$th agent, 
 interacting with the $j$th agent (chosen randomly from one of the four nearest neighbors of the $i$th agent),
the opinion value $o_i$  changes according to
\begin{equation}
  o_i(t+1)=o_i(t)+\mu_{ij}o_j.
\end{equation}
No sum over the index $j$ is implied, as the model is binary-exchange.  A non-linearity enters the model from the
imposed bounds in the opinion values for the extreme ends at $\pm 1$, i.e., $|o_i|\le 1$, signifying the limit to an extreme opinion.
$\mu_{ij}$ is an annealed variable assuming the values $\pm 1$, and is negative with probability $p$. An order disorder transition occurs close to $p_c \approx 0.11$ here \cite{nuno2,sudip}. 

\begin{figure}
\includegraphics[width = 14cm]{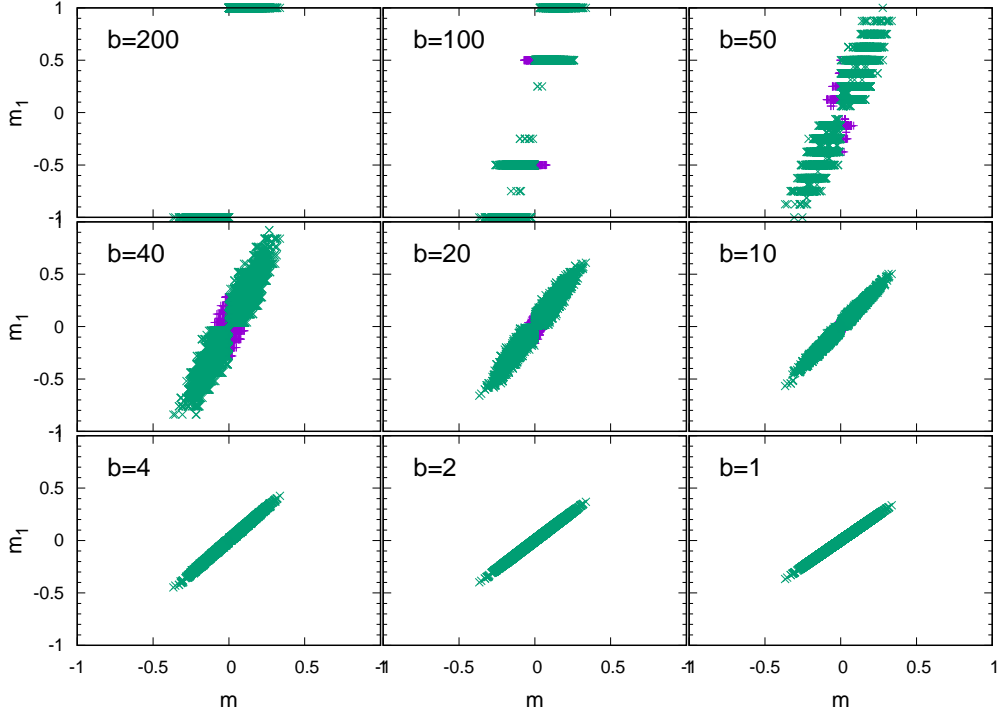}
\caption{In the Ising model near critical point, the magnetization of the original lattice ($m$) is plotted against that of the course grained lattice ($m_1$) for different values of $b$, the coarse graining box length. The cases where the signs of these two quantities are different from each other (second and fourth quadrant; points marked in  purple), the minority candidate wins. Clearly, this is not possible in the limits $b=1$ and $b=L$ (here $L=200$). However, in other cases such situations exist (see text for further quantification).} 
\label{major_minor} 
\end{figure}

The normalized order parameter $m$  in the two models are given by $\sum_i s_i/N$ and $\sum_i o_i/N$ where $N$ is the total number of agents; 
$N = L^2$ in a $L \times L$ lattice. 
The initial state is chosen to be totally random for both models corresponding to a high noise value. 
The system gradually relaxes towards equilibrium state, which is still disordered but having large spatial correlation.  
The order parameter  oscillates about zero close to the critical point as a 
function of time. For each time step, if the order parameter $m$ is not exactly 
zero, the coarse graining is applied with different scale factors $b$. The magnetization $m_1$,  after  coarse graining,  is calculated. We estimate the 
probability $\epsi (b,L)$  that  $mm_1 < 0$. The variation of this probability with $b$ for different values of $L$ are then studied. 

%The whole process can also be regarded as one of information transmission. 
%If $MM^\prime < 0$, the original information is lost and hence the 
%probability that $MM^\prime < 0$ is called the error $E$.

Another quantity that is estimated is how the error $\epsi$  propagates in successive coarse graining. For this we have considered a two step procedure.  If the one step coarse graining is done with scale factor $b$, we now consider successive 
coarse graining procedures by factors $b_1$ and $b_2$ respectively 
%representing the first and 
%second respectively 
with $b_1b_2 = b$. It is of course well known in 
critical phenomena that the the two procedures lead to the same behavior as far as normalization of the parameters are concerned 
\cite{skma}. 
However, the question we ask here is somewhat different. 
In an election procedure, the winner is usually decided on a majority 
vote basis. 
Subsequently, the issues to be   resolved  
are  decided on the basis of the votes 
by the elected representatives. 
The outcome, however, could be different  
if all the individuals were allowed to vote on these issues. 
So the  two step procedure measures how far the people's decision 
could be  propagated to a higher level. 
In the context of information theory, it could be regarded as a two step coding. 
%This is also an important problem in information theory ({\color {red} Soumya})
%In terms of information processing, it measures the propagation of the error that is involved in a two step 
%process compared to the one step case.
%It is also possible to factorise $b$ in different ways and again we study how the error is different in this case. 

In the simulations on the square lattices, periodic boundary condition has been 
 used and  several configurations are run to get the average values.

\section{Results} 
\subsection{One step coarse graining}
As mentioned before, the electoral college system in the US presidential election resembles a single step coarse graining process, with each coarse graining block 
representing a state. 
Here we report the results of a single step coarse graining when the number of coarse graining blocks covering the system is varied i.e., the block sizes are varied. 
In Fig. \ref{major_minor} the magnetization of the original lattice is plotted against that of the coarse grained lattice for the Ising model. For different values of $b$,
the points where the signs of these two quantities are different, varies. In the extreme limits $b=1$ and $b=L$ such points do not exist. We now go on to further quantify the variations in probabilities of such cases.

\begin{figure}
\includegraphics[width = 8cm]{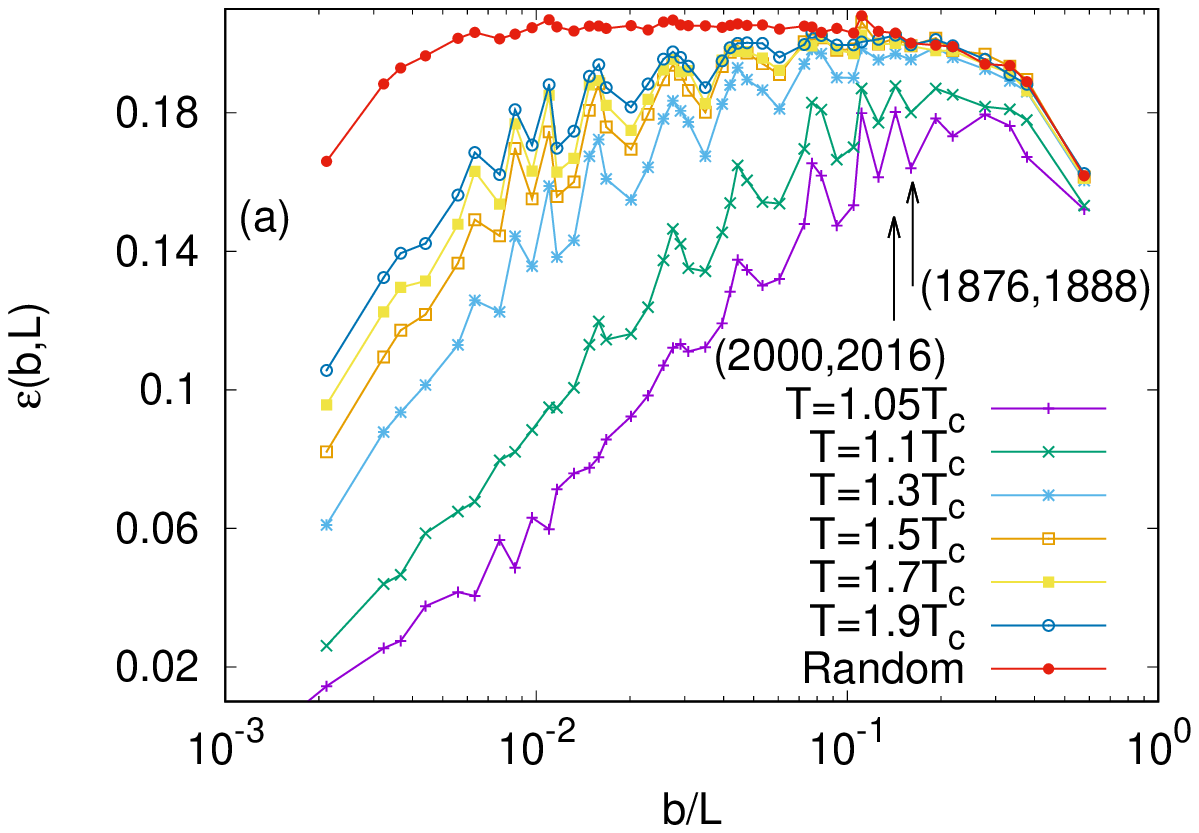}
\includegraphics[width = 8cm]{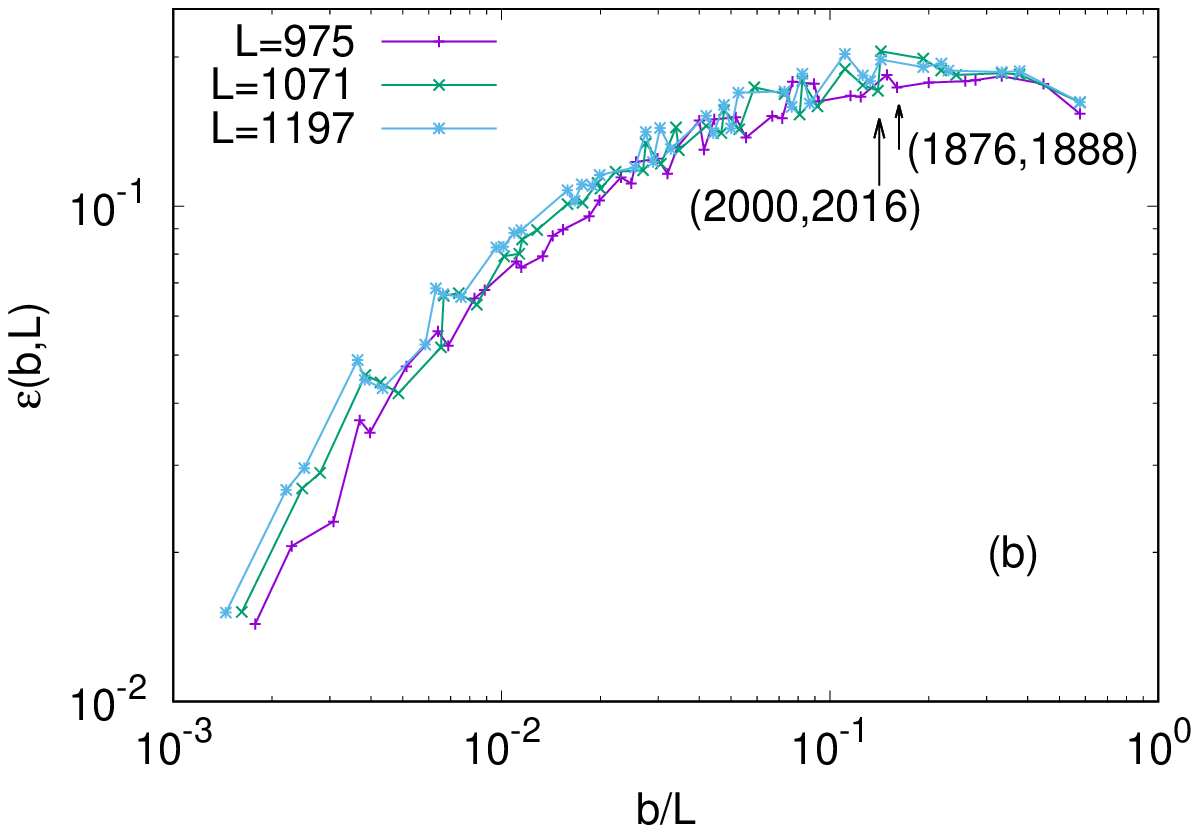}
\caption{Variation of the minority win probability $\epsilon(b,L)$ for the Ising model.  (a) The variation of the minority win probability $\epsilon(b,L)$ is shown for the Ising model in two dimensions for $L=819$ at different temperatures above $T_c$. As the temperature is raised significantly higher than $T_c$, the curves approach the random configuration limit, as expected. The arrows indicate the points where the events of minority win did occur in the past. (b) The minority win probabilities are plotted for different system sizes near the critical point. A data collapse is seen for the scaling variable $b/L$. As before, the arrows indicate the events of minority win. It is clear that for further rise in the number of states would significantly reduce the minority win probability.}
\label{ising_fig} 
\end{figure}

\begin{figure}
\includegraphics[width = 8cm]{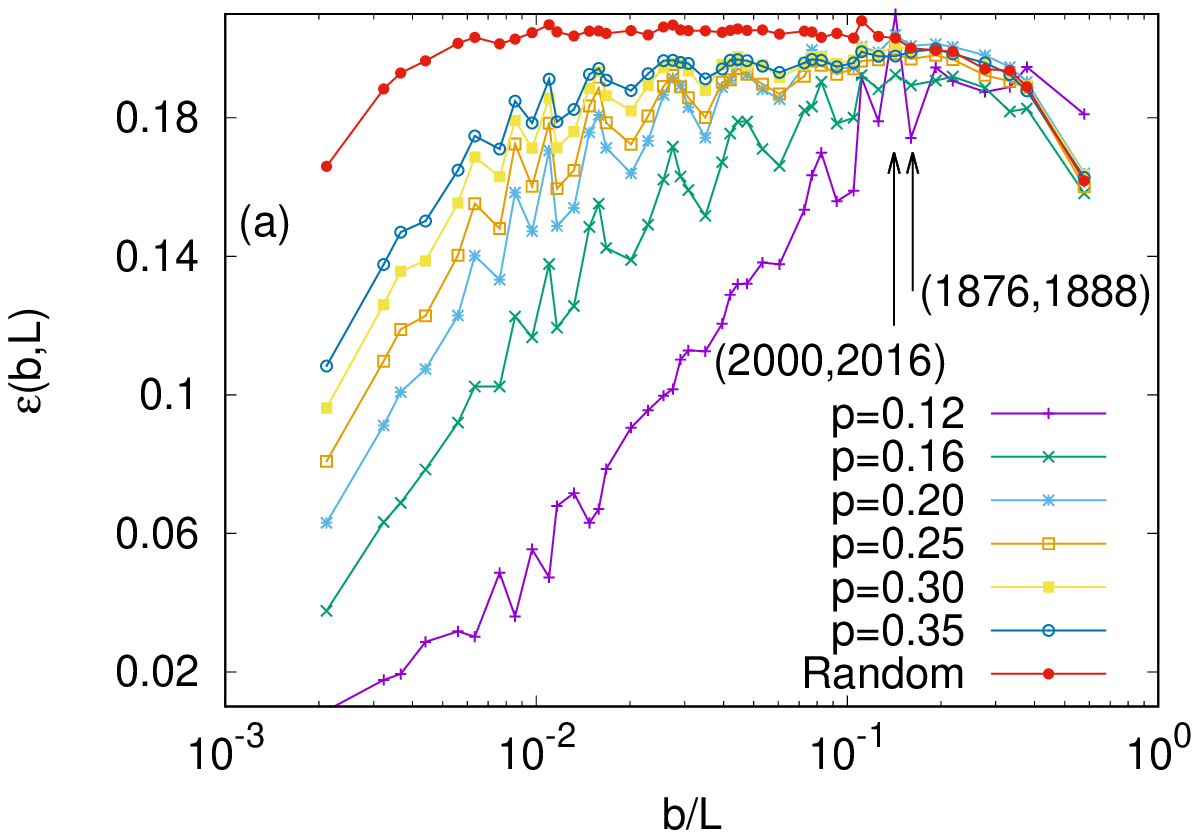}
\includegraphics[width = 8cm]{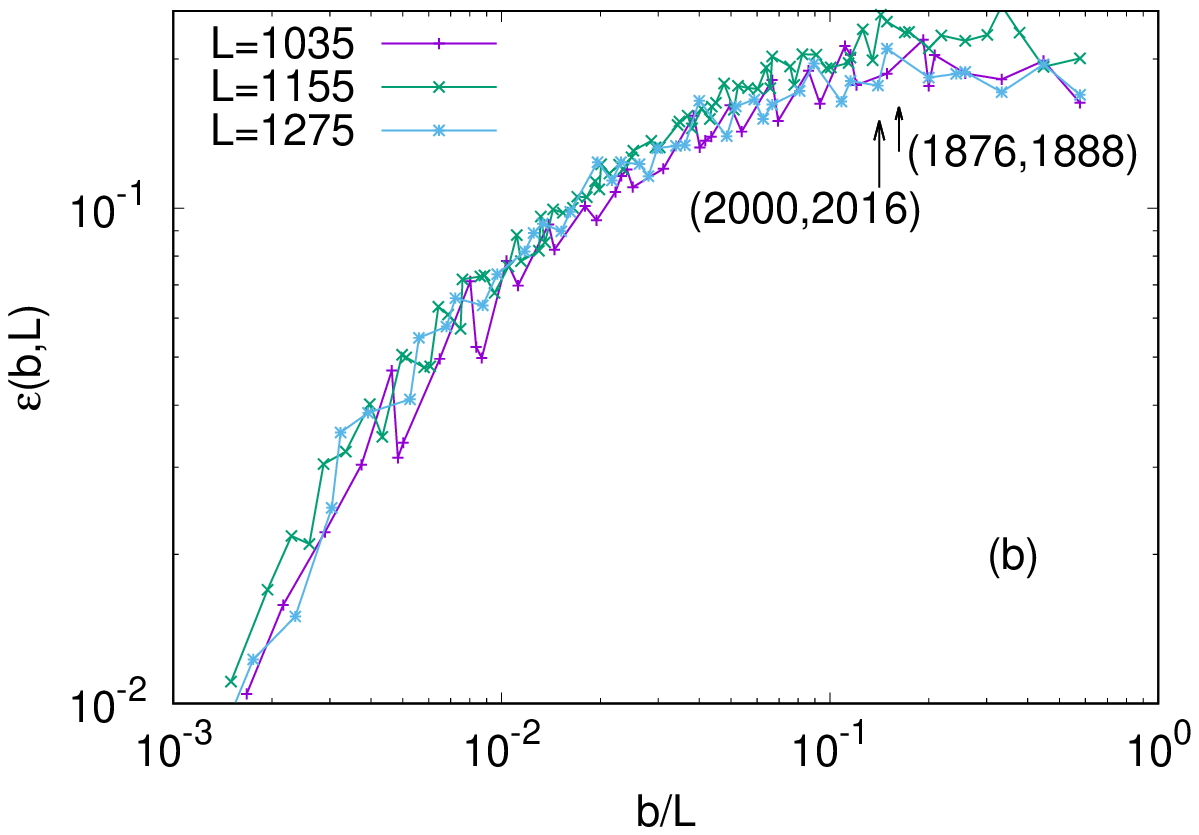}
\caption{The variation of the minority win probability $\epsilon(b,L)$ of the KEM. (a) The variation of the minority win probability $\epsilon(b,L)$ is shown for the KEM in two dimensions for $L=819$ at different values for the parameter $p$ above $p_c$. As the parameter $p$ is raised significantly higher than $p_c$, the curves approach the random configuration limit, as expected. The arrows indicate the points where the events of minority win did occur in the past. (b) The minority win probabilities are plotted for different system sizes near the critical point. A data collapse is seen for the scaling variable $b/L$. As before, the arrows indicate the events of minority win. It is clear that for further rise in the number of states would significantly reduce the minority win probability.}
\label{kem_fig} 
\end{figure}

Although we are interested in the case when the system is close to criticality,
one can calculate the errors for higher noise values as well. 
%For a fixed $b$ value it was shown in \cite{} that the error becomes larger above criticality and approaches  the results of a random case. 
We calculate $\epsi(b,L)$ for a fixed value of $L$ 
for different noise values for the Ising model and the KEM,
shown in Fig. \ref{ising_fig}a and \ref{kem_fig}a.
As expected, $\epsi$ increases with the noise factor. 
We also note that $\epsi$, which is zero at the extreme limits, 
increases with $b$ and shows a shallow peak at $b \sim b_p$ only very close to the maximum value of $b = L/\sqrt{2}$. 
%We find that clearly 
%such that even 
For larger values of the noise, the results are almost independent of $b$
as the system approaches  a random case, also shown for comparison.

Next, the results for different system sizes very close to  criticality are
presented in Figs \ref{ising_fig}b and \ref{kem_fig}b where $\epsi$  
is  plotted for different system sizes. We note that the 
data collapse when they are plotted  against $b/L$. 
It may be added that the same happens for the random case, and therefore presumably for all values of the noise factor above criticality. 

Hence $b/L$ is identified as the scaling variable. We note that the scaling collapse becomes better as the system sizes are increased, in general there are quite strong finite 
size effects. Specifically, 
the error $\epsilon(b,L)$ decreases for smaller system sizes. Intuitively it is clear, as the effective `critical point' of smaller systems 
(say, the peak of the heat capacity in the Ising model) is higher than that of the thermodynamic limit, the order parameter at a given distance 
from the thermodynamic critical point will increase with the decrease in the system size and hence will give lower values of $\epsilon$. 
%{\color {blue} should we mention about strong finite size effects?})

Now, clearly the variable $b/L=1/\sqrt{M}$, where $M$ is the number of states. In the case of the US, as shown in Fig. \ref{states}, 
the number of states in the US has varied considerably over the years. Particularly, the quantity $1/\sqrt{M}$ has changed between 
$1/\sqrt{11}\approx 0.301$ in 1788 to $1/\sqrt{50}\approx 0.141$ from 1959 onward. Comparing these values with the plots in
Figs \ref{ising_fig}b and \ref{kem_fig}b, we see that the probability of PC losing has changed with time and currently it is very close to the
point beyond which the probability will decrease drastically if the number of states in increased, i.e., $b$ decreased in the coarse graining. In the Figs \ref{ising_fig} and \ref{kem_fig},
the arrows indicate the years of PC losing.

\subsection{Two step coarse graining}

We now discuss  the two step 
 coarse graining done with scale factors (block size) $b_1$ and $b_2$ and compare the results  with the one step process with $b=b_1b_2$. We also compare the 
results when the order of the two step process is reversed, i.e., first with a scale factor $b_2$ and then $b_1$. We use the convention $b_2 > b_1$.

%for both the Ising model and the KEM near criticality, when $b_1\ne b_2$ the order in which the coarse graining operations are performed 
%change the final error $\epsilon(b_1,b_2,L)$ systematically.

We first consider the random case which corresponds to infinite noise.
Here it is found that the two step process significantly increases the 
error  but the order hardly matters. 
Table \ref{table1} shows the errors in the two step process. 
%One can observe that indeed, the error increases significantly. 
%we have also considered the  two step procedure with the coarse graining with $b_2$ first and then $b_1$ to compare the results. 

%The results are presented for the random case, the IM and the KEM. 

Let  $\epsi^\prime (b_1,b_2,L)$ denote the
error for the two step process, i.e., at the end of the two coarse grainings.
%$b_1b_2 = b$ where $b$ is the one step case.
After the first step, the fraction which retains the original sign of the 
order parameter is
$(1-\epsi(b_1,L))$. 
The probability  the  sign is changed in the second step  
 for this fraction is  $\epsi(b_2,L/b_1)$.
For the fraction 
$\epsi(b_1,L)$ 
for  which the  sign did  change  after the first step,   
the contribution to $\epsi^\prime$ will come if these configurations  retain the signature
in the second step. This happens
with probability 
$(1-\epsi(b_2,L/b_1))$.
Hence 
\be 
\epsi^\prime(b_1,b_2,L) = \epsi(b_2,L/b_1)(1-\epsi(b_1,L)) +
(1-\epsi(b_2,L/b_1))\epsi(b_1,L).
\ee
In the scaling regime, $\epsi (b,L) $ is a function of  $b/L$ only.
So
\begin{equation}
 \epsi^\prime(b_1, b_2, L) = \epsi(b/L)(1-\epsi(b_1/L)) + (1-\epsi(b/L))\epsi(b_1/L).
\label{eqepsiprime}
\end{equation}
Let us call $\epsi(b/L) = y$ and $\epsi(b_1/L) = x$.
Then the RHS of above eq becomes:
\begin{equation}
\label{xy}
y(1-2x) +x
\end{equation}

This quantity will increase with $x$ unless $y$ is greater than 0.5 which 
is usually not the case. 
One can proceed further for the random case
and  obtain an upper bound for the error.
Here, we have noted that $\epsi$ has a monotonic  slow increase 
for small values  
 of $b$  and attains a constant value for larger $b$ values. 
So an upper bound on $x$  is $y$ (since $b_1 < b$). 
 $y$ has  more or less a constant value $\approx 0.2$ (numerically
obtained)    
 and therefore $\epsi^\prime
\leq 2y(1-y) \approx 0.32$. 
%for $b$ large but $< L/\sqrt{2}$ (see table for numerically obtained values),
%the upper bound is $\approx 0.32$
We indeed get  that the two step procedure gives values less than this upper bound (see Table \ref{table1}).

According  to Eq. (\ref{eqepsiprime}), the results should depend on the order in which
the 2 step coarse graining is done. For the reverse order we will get Eq. (\ref{xy}) where
$x$ is replaced by $x' = \epsi(b_2/L)$. 
However, for the random case, as $\epsi$ remains almost constant  for a considerable range of values of   
 $b/L$ (see Figs \ref{ising_fig}, \ref{kem_fig}),   we get negligible difference when the order is changed.

%Are the results coming out to be indep of order as the Random case is being considered? SB said he has a program for KEM. Can one or or two cases be simulated for comparison?}}
%
  
On the other hand,  for both KEM and IM, close to criticality, 
we find that  there is an appreciable 
dependence of $\epsi$ on $b/L$ and the results for the two step process are found to be  sensitive to the order (see Tables \ref{table2} and \ref{table3}). 
In particular, we note that when the coarse graining is done with $b_1 < b_2$ first, the resultant errors are more or less same as that for the two step process. This might not be expected from eq. \ref{xy}, however, it must be remembered that
the above analysis is made  for the thermodynamic limit, $L \to \infty$; 
the finite size effects  mentioned earlier  will be enhanced  in a two step process.

%in finite systems  the dependence of the critical temperature is crucial as 
%mentioned before. 

When  $b_2$ is used first, the error increases. This is not difficult to explain: a larger value of $b$ gives larger errors after the first step  (unless it is beyond $b_p$ where a peak value occurs which has not been considered). 
Naturally,  in the second step, when the system size has got reduced,  even
with $b_1$ smaller, the error is increased.  

\begin{table}
\centering
\caption{Two step coarse graining in the random case}
\label{table1}
\begin{tabular}{|l|l|l|l|l|l|}
\hline
$L$ & $b_1$,$b_2$ & $b_1b_2/L$& $\epsilon^\prime(b_1,b_2,L)$ & $\epsilon^\prime(b_2,b_1,L)$ & $\epsilon(b_1 b_2,L)$\\
\hline
\hline
495 & 3,5 & 0.030 & 0.269 & 0.270 & 0.205\\
\hline
495 & 3,15 & 0.091 & 0.272 & 0.272 & 0.204\\
\hline
495 & 5,9 & 0.091 & 0.277 & 0.276 & 0.204\\
\hline
495 & 3,55 & 0.333 & 0.262 & 0.264 & 0.191\\
\hline
495 & 5,33 & 0.333 & 0.268 & 0.270 & 0.191\\
\hline
495 & 11,15 & 0.333 & 0.271 & 0.270 & 0.191\\
\hline
\hline
585 & 3,5 & 0.026 & 0.268 & 0.268 & 0.203\\
\hline
585 & 3,39 & 0.200 & 0.268 & 0.270 & 0.201\\
\hline
585 & 9,13 & 0.200 & 0.277 & 0.277 & 0.201\\
\hline 
585 & 3,65 & 0.333 & 0.263 & 0.265 & 0.193\\
\hline 
585 & 5,39 & 0.333 & 0.269 & 0.269 & 0.193\\
\hline 
585 & 13,15 & 0.333 & 0.274 & 0.272 & 0.193\\
\hline
\hline
693 & 3,7 & 0.030 & 0.273 & 0.269 & 0.204\\
\hline
693 & 3,33 & 0.143 & 0.273 & 0.274 & 0.204\\
\hline
693 & 9,11 & 0.143 & 0.276 & 0.276 & 0.204\\
\hline
693 & 3,77 & 0.333 & 0.266 & 0.264 & 0.194\\
\hline
693 & 7,33 & 0.333 & 0.270 & 0.272 & 0.194\\
\hline
693 & 11,21 & 0.333 & 0.270 & 0.272 & 0.194\\
\hline

\end{tabular}
\end{table}

\begin{table}
\centering
\caption{Two step coarse graining in the Ising model}
\label{table2}
\begin{tabular}{|l|l|l|l|l|l|}
\hline
$L$ & $b_1$,$b_2$ & $b_1 b_2/L$& $\epsilon^\prime(b_1,b_2,L)$ & $\epsilon^\prime(b_2,b_1,L)$ & $\epsilon(b_1 b_2,L)$\\
\hline
\hline
495 & 3,5 & 0.030 & 0.086 & 0.091 & 0.084\\
\hline
495 & 3,15 & 0.091 & 0.145 & 0.165 & 0.142\\
\hline
495 & 5,9 & 0.091 & 0.149 & 0.154 & 0.142\\
\hline
495 & 3,55 & 0.333 & 0.177 & 0.222 & 0.175\\
\hline
495 & 5,33 & 0.333 & 0.179 & 0.214 & 0.175\\
\hline
495 & 11,15 & 0.333 & 0.188 & 0.195 & 0.175\\
\hline
\hline
585 & 3,5 & 0.026 & 0.096 & 0.099 & 0.093\\
\hline
585 & 3,39 & 0.200 & 0.186 & 0.236 & 0.185\\
\hline
585 & 9,13 & 0.200 & 0.195 & 0.201 & 0.185\\
\hline
585 & 3,65 & 0.333 & 0.187 & 0.238 & 0.186\\
\hline
585 & 5,39 & 0.333 & 0.191 & 0.233 & 0.186\\
\hline
585 & 13,15 & 0.333 & 0.203 & 0.209 & 0.186\\
\hline
\hline
693 & 3,7 & 0.030 & 0.112 & 0.118 & 0.109\\
\hline
693 & 3,33 & 0.143 & 0.180 & 0.218 & 0.178\\
\hline
693 & 9,11 & 0.143 & 0.188 & 0.192 & 0.178\\
\hline
693 & 3,77 & 0.333 & 0.181 & 0.241 & 0.179\\
\hline
693 & 7,33 & 0.333 & 0.186 & 0.222 & 0.179\\
\hline
693 & 11,21 & 0.333 & 0.193 & 0.213 & 0.179\\
\hline
\hline

\end{tabular}
\end{table}

\begin{table}
\centering
\caption{Two step coarse graining in the kinetic exchange model}
\label{table3}
\begin{tabular}{|l|l|l|l|l|l|}
\hline
$L$ & $b_1$,$b_2$ & $b_1b_2/L$& $\epsilon^\prime(b_1,b_2,L)$ & $\epsilon^\prime(b_2,b_1,L)$ & $\epsilon(b_1 b_2,L)$\\
\hline
\hline
495 & 3,5 & 0.030 & 0.063 & 0.065 & 0.063\\
\hline
495 & 3,15 & 0.091 & 0.118 & 0.138 & 0.121\\
\hline
495 & 5,9 & 0.091 & 0.117 & 0.122 & 0.121\\
\hline
495 & 3,55 & 0.333 & 0.169 & 0.219 & 0.169\\
\hline
495 & 5,33 & 0.333 & 0.168 & 0.203 & 0.169\\
\hline
495 & 11,15 & 0.333 & 0.168 & 0.176 & 0.169\\
\hline
\hline
585 & 3,5 & 0.026 & 0.0793 & 0.081 & 0.082\\
\hline
585 & 3,39 & 0.200 & 0.181 & 0.228 & 0.179\\
\hline
585 & 9,13 & 0.200 & 0.178 & 0.194 & 0.179\\
\hline
585 & 3,65 & 0.333 & 0.170 & 0.238 & 0.169\\
\hline
585 & 5,39 & 0.333 & 0.170 & 0.219 & 0.169\\
\hline
585 & 13,15 & 0.333 & 0.177 & 0.184 & 0.169\\
\hline
\hline
693 & 3,7 & 0.030 & 0.090 & 0.097 & 0.091\\
\hline
693 & 3,33 & 0.143 & 0.190 & 0.210 & 0.192\\
\hline
693 & 9,11 & 0.143 & 0.185 & 0.188 & 0.192\\
\hline
693 & 3,77 & 0.333 & 0.183 & 0.234 & 0.183\\
\hline
693 & 7,33 & 0.333 & 0.183 & 0.215 & 0.183\\
\hline
693 & 11,21 & 0.333 & 0.187 & 0.204 & 0.183\\
\hline
\hline

\end{tabular}
\end{table}

\section{Discussion and conclusions}
The indirect nature of the election of the US president highlights the importance of the coarse graining process
while modeling the voting process using discrete Ising-symmetric models.
The process of coarse-graining, particularly near the critical point of a spin system, is supposed to keep the system invariant under a renormalization group sense. However, as was noted before (see e.g. Ref. \cite{bcs_us}), the `invariance' does not guarantee that the sign of the magnetization of the original and the coarse grained lattice would remain the same. In fact, the probability of such events is known to vary systematically, showing finite size scaling behavior, near the critical point.
Indeed, the process of coarse-graining is a loss of information that can have very significant effect where the final sign of the order parameter matters. One such situations is the US presidential election. The electoral college system in the US, that assigns all delegates of the winning candidate in a state, is similar to a process of coarse-graining. In this context, a difference in the sign of the magnetization in the original and coarse grained lattice would mean that the candidate winning most of the popular votes did not win the overall election.

 So far as the dissimilarity of the final
outcome of the election result and the popular vote is concerned, it is expected that the effect will be most 
relevant when the elections are closely contested and there is spatial fluctuation in the voting pattern.
In that case, if the votes of one candidate is heavily concentrated in a few states, while the other candidate wins
in more number of states even though marginally, the latter would win the election due to the effective single step coarse graining 
coming from the electoral college system.

 This then
motivates the quantification of the probability of the minority win $\epsilon(b,L)$ as a function of the system size $L$ and the coarse graining
block size $b$, in the models such as the
Ising and the KEM. Interestingly, $b/L$ emerges as the relevant scaling variable, at least in the limit of the 
large system sizes (see Fig. \ref{ising_fig}, \ref{kem_fig}). However, it is obvious that in the two extreme limits $b=1$ and $b=L$, the result of the 
coarse graining has no significance and $\epsilon$ is exactly zero (see Fig. \ref{major_minor}). In the intermediate range, therefore, $\epsilon(b/L)$ will
show a point of maximum. It is not obvious at which point the maximum would occur, but given that $b/L$ is the scaling variable, 
it is determined by the critical fluctuation of the model and not the system or block sizes. 

Now, given that the number of states in the US has varied over the years and that the minority win probability $\epsilon(b/L)$
has a non-monotonic variation, it is interesting to check according to the prediction of this model, how has $\epsilon(b/L)$ 
changed over the years for the US presidential election. Indeed, as is indicated in Fig. \ref{states}, the effective $b/L$ values, estimated 
as $1/\sqrt{M}$ ($M$ is the number of states), is such that $\epsilon(b/L)$ has been close to its maximum. In fact, it can also be 
seen from Figs \ref{ising_fig} and \ref{kem_fig} that for larger values of $M$ i.e., by splitting up larger states, the minority win probability can be sharply reduced. 
However, it should also be mentioned here that it is an idealized situation and in practice there are some mechanisms in place to counter 
the effect introduced by the electoral college viz., the variation in the number of delegates according to the sizes of the states, which we did not
consider here to keep things simple. 

Finally, if the coarse graining process is repeated a second time, the scale factors being $b_1$ and $b_2$ in the 
first and second steps respectively, the errors $\epsilon^\prime (b_1,b_2,L)$ are  higher
for the random cases. Here, one can estimate an upper bound based on the 
numerical results. For the two models used here, the error depends on the order in which the two step process is implemented. 
In general, as can be seen from the tables, systematically $\epsilon^\prime (b_2,b_1,L) >\epsilon^\prime (b_1,b_2,L)$ if $b_2>b_1$ near the critical points of the
models in two dimensions. The results when the smaller of $b_1,b_2$ is taken first in fact yields an 
error not much different from the one step case with $b= b_1b_2$. This,  however, is not true for random cases (or when the noise is far above the critical value), where the 
coarse graining processes commute. Therefore, this 
observation can also be attributed to the critical fluctuations of the model. The two step coarse graining is relevant in the real world situations
where the elected members of a legislative body can further form coalitions among themselves. $\epsi^\prime $ measures the chances of their decisions not 
being aligned with that of their electorates. It is also relevant for coding in information theory,  the error 
is expected to increase if the coding is done in two steps. 

The above mentioned results remain qualitatively true in the mean field limit i.e., a fully connected topology (see also \cite{bcs_us}). However, a more realistic 
topology for the interaction domains of the agents would be a network structure that can more closely resembles social connectivity \cite{ba_rmp}. Furthermore, 
the assumptions of equal sizes for each coarse grainin box (each state) could be made more realistic and the implicit assumption of equal population densities 
in the states could also be varied according to data. 

In conclusion, we have reported the effect of coarse graining in the elections where an intermediate body is present between the 
population and the winner, for example in the US presidential election. We showed, using finite size scaling of the Ising model and
kinetic exchange opinion models near criticality that the probability of a candidate winning the election without winning the popular 
vote depends non-monotonically with the coarse graining block size. Furthermore, using the data for the number of states in the US, 
we show that according to the model studied here, the minority win probability is near to the maximum value and could sharply decrease, provided
the number of states are increased further.

Acknowledgments: PS is grateful to the late Dietrich Stauffer who inspired novel research ideas and application of statistical
physics in social phenomena.
Financial support from SERB scheme 
EMR/2016/005429 (Government of India) is also acknowledged.

%%%%%%%%%%%%%%%%%%%%%%%%%%%%%%%%%%%%%%%%%%%%%%%%%%%%%%%%%%%%%%%%%%%%%%%%%%%%%%%%%


\begin{thebibliography}{99}
\bibitem{stauffer} 
D. Stauffer, {\it Opinion dynamics and sociophysics}, in: R. Meyers (eds) Encyclopedia of Complexity and Systems Science. Springer, New York, NY (2009).

\bibitem{sen_chak}
P. Sen, B. K. Chakrabarti, {\it Sociophysics: An Introduction}, Oxford University Press, Oxford (2014).

\bibitem{soc_rmp}
C. Castellano, S. Fortunato, V. Loreto, {\it Statistical physics of social dynamics}, Rev. Mod. Phys. {\bf 81},
591 (2009).

\bibitem{galam_book}
S. Galam, {\it Sociophysics: A Physicist's Modeling of Psycho-political Phenomena}, Springer, Boston, MA (2012).

\bibitem{cliff}
P. Clifford, A. Sudbury, {\it A model for spatial conflict}, Biometrika {\bf 60}, 581 (1973).

\bibitem{ligg1}
T, M. Liggett, {\it Interacting particle systems}, Springer, New York, NY (1985).

\bibitem{ligg2}
T. M. Liggett, {\it Interacting particle systems: Contact, voter and exclusion processes}, Springer-Verlag, Berlin (1999).

\bibitem{bcs_us}
S. Biswas, P. Sen, {\it Critical noise can make the minority candidate win: The US presidential election cases}, Phys. Rev. E {\bf 96}, 032303 (2017).

\bibitem{bcs_brexit} S. Mukherjee, S. Biswas and P. Sen, {\it  Long route to consensus: Two stage coarsening in binary choice voting
    model}, Phys. Rev. E
{\bf 102}, 012316 (2020).

\bibitem{sg1} S. Galam, {\it Social paradoxes of majority rule voting and renormalization group},
J. Stat. Phys. {\bf 61}, 943 (1990).


\bibitem{sg2}
S. Galam, {\it Geometric vulnerability of democratic institutions against 
lobbying: A sociophysics approach}, Math. Models Methods Appl. Sci. {\bf 27}, 13 (2017).


\bibitem{erikson}
R. S. Erikson, K. Sigman, L. Yao, {\it Electoral college bias and the 2020 presidential election}, PNAS {\bf 117}, 27940 (2020).

\bibitem{election_book}
Paul F. Kisak (ed.), {\it The U. S. Presidential election process}, CreateSpace Independent Publishing Platform, 2016. 

\bibitem{skma} See e.g,, S. K.  Ma, {\it Modern theory of critical phenomena}, Taylor and Francis, New York (1976).

\bibitem{bcs}
S. Biswas, A. Chaterjee, P. Sen, {\it Disorder induced phase transition in kinetic models of opinion formation},
Physica A {\bf 391}, 3257 (2012).

\bibitem{nuno2}
N. Crokidakis, {\it Phase transition in kinetic exchange opinion models
with independence}, Phys. Lett. A {\bf 378}, 1683 (2014).

\bibitem{sudip}
S. Mukherjee, A. Chatterjee, {\it Disorder-induced phase transition in an opinion dynamics model:
Results in two and three dimensions}, Phys. Rev. E {\bf 94}, 062317 (2016).

\bibitem{ba_rmp}
R. Albert, A. Barabasi, {\it Statistical mechanics of complex networks}, Rev. Mod. Phys. {\bf 74}, 47 (2002).


\end{thebibliography}
\end{document}